\documentclass[11pt,aps,tightenlines]{revtex4-2}

\usepackage{amsmath}
\usepackage{amssymb}
\usepackage{amsfonts}
\usepackage{bm}
\usepackage[mathscr]{eucal}
\usepackage{theorem}
\usepackage{psfrag}
\usepackage{subfigure}
\usepackage{wasysym}
\include{graphix}
\usepackage{framed}
\usepackage{enumitem}%
\usepackage{lipsum}
\usepackage{float}
\usepackage{MnSymbol}

\usepackage{tikz}

\usetikzlibrary{patterns}
\usetikzlibrary{shapes}
\def\normaltwo{\x,{0.7+1/exp(((\x-6)^2)/2)}}

\newcommand{\mathd}{\mathrm{d}}
\newcommand{\mathe}{\mathrm{e}}

\newcommand{\myRe}{\mathrm{Re}}
\newcommand{\myWe}{\mathrm{We}}

\newcommand{\maxR}{\mathcal{R}_{max}}

\newcommand{\mysmall}{\varepsilon }

\newcommand{\thetadyn}{\vartheta_{a}}

\newcommand{\myqed}{\nobreak \ifvmode \relax \else
      \ifdim\lastskip<1.5em \hskip-\lastskip
      \hskip1.5em plus0em minus0.5em \fi \nobreak
      \vrule height0.75em width0.5em depth0.25em\fi}

\begin{document}

\title{Bounds on the Spreading Radius in Droplet Impact: The Inviscid Case}

\author{
Alidad Amirfazli$^\text{1}$, Miguel D. Bustamante$^\text{2}$, Yating Hu$^\text{1}$, Lennon \'O N\'araigh$^\text{2}$}

\affiliation{$^\text{1}$Department of Mechanical Engineering, York University, Toronto, Ontario M3J 1P3, Canada\\
$^\text{2}$School of Mathematics and Statistics, University College Dublin, Belfield, Dublin 4, Ireland
}




\begin{abstract}
We consider the classical problem of droplet impact and droplet spread on a smooth surface in the case of an ideal inviscid fluid.  We revisit the rim-lamella model of Roisman et al. [\textit{Proceedings of the Royal Society of London. Series A: Mathematical, Physical and Engineering Sciences}, 458(2022), pp.1411-1430.].  This model comprises a system of ordinary differential equations (ODEs); we present a rigorous theoretical analysis of these ODEs, and derive upper and lower bounds for the maximum spreading radius.  Both bounds possess a $\mathrm{We}^{1/2}$ scaling behaviour, and by a sandwich result, the spreading radius itself also possesses this scaling.  We demonstrate rigorously that  the rim-lamella model is self-consistent: once a rim forms, its height will invariably exceed that of the lamella.  We introduce a rational procedure to obtain initial conditions for the rim-lamella model.  Our approach to solving the rim-lamella model gives predictions for the maximum droplet spread that are in close agreement with existing experimental studies and direct numerical simulations.
\end{abstract}

\maketitle

\section{Introduction}
\label{sec:intro}

The impact of a droplet on a smooth, homogeneous substrate is a well studied problem, with many practical applications, e.g. inkjet printing~\cite{yarin2006drop}, cooling~\cite{yarin2006drop,valluri2015}, and crop spraying~\cite{yarin2006drop,Moghtadernejad2020}.
Droplet-impact regimes are studied  in terms of the droplet's Weber number ($\myWe$) and Reynolds number ($\myRe$).  For instance, there is a splash parameter $\myWe\sqrt{\myRe}$ which determines a threshold above which splash occurs~\cite{mundo1995droplet,josserand2016drop}.  The threshold value is not universal~\cite{marengo2011drop}, and different experiments have produced different values, a summary of which is provided in~\cite{moreira2010advances}.  Nevertheless, the threshold is of the order of $10^3$--$10^4$.

\noindent   Just below this threshold, and typically for $\myWe\geq 10^2$ and $\myRe\geq 10^3$~\cite{de2010thickness}, there is `rim-lamella' regime, in which the droplet flattens and spreads into an axisymmetric structure involving a lamella or sheet, with a much thicker rim forming at the sheet extremity. 
The maximum extent of spreading of the rim-lamella structure is of interest due to its effect on the aforementioned applications.    The maximum spreading radius is denoted by $\maxR$, and is a function of  $\myWe$ and   $\myRe$.  Here, and throughout the present work, we take:
	\begin{equation}
	\myWe=\frac{\rho U_0^2 R_0}{\sigma},\qquad \mathrm{Re}=\frac{\rho U_0 R_0}{\mu},
	\label{eq:Wedef}
	\end{equation}
	where $\rho$ is the liquid density,   $\mu$ is the liquid viscosity, and $\sigma$ the surface tension.  Also, $U_0$ is the droplet's speed prior to impact and $R_0$ is droplet radius prior to impact. 

\subsection{Aim of the Paper}

There is no exact formula for $\maxR(\myWe,\myRe)$. In the inviscid limit, dimensional analysis or an energy-budget analysis demonstrates that $\maxR\sim \myWe^{1/2}$.  The aim of the present work is to derive this scaling law rigorously.   Although the scaling behaviour is already well known, it is worthwhile to explore its origin. For this purpose, we use an established rim-lamella model, based on a set of ordinary differential equations.  The point of departure is the theoretical analysis of the differential equations:  we use  estimates based on differential inequalities (including Gronwall's inequality) as well as perturbation theory to prove rigorously the scaling behaviour $\maxR\sim \myWe^{1/2}$.
A second aim is to show that such rim-lamella models are self-consistent, in the sense described below. 

We are motivated to pursue this approach as a novel  application of the theory of estimates.  
The use of estimates in Fluid Mechanics is particularly fruitful, as it yields rigorous results for highly complex systems of systems of equations, albeit that the results are in terms of inequalities.  To date, the theory  has been used to put constraints on the critical Reynolds number in nonlinear instability of channel flow~\cite{doering1995applied}, the expected regularity of the solutions of the Navier--Stokes equations~\cite{doering1995applied}, quantify mixing efficiency~\cite{thiffeault2004bound}, and identify the conditions under which thin films will rupture~\cite{naraigh2010nonlinear}.
The application of estimates to rim-lamella models (not only in the inviscid case, as in the present article, but later also to the viscous case) can provide fresh understanding of droplet impact and droplet spreading dynamics.

\subsection{Literature Review}

Different approaches to the computation of the maximum spreading radius exist, including an energy-budget analysis (e.g.~\cite{chandra1991collision}).  This  relates the pre-impact energy of the droplet to the post-impact energy at maximum spreading, of which the latter can be assumed to be entirely due to surface energy.  By treating the droplet at maximum spreading  as having a pancake-like structure, a simple closed expression can be found for $\maxR$ as a function of $\mathrm{We}$.  Viscous dissipation  can be incorporated into the energy budget by explicitly modelling the boundary layer in the lamella~\cite{wildeman2016spreading}.  This then gives an estimate for $\maxR$ as a function both of $\mathrm{We}$, and $\mathrm{Re}$.
This simplified approach omits the rim.  The approach has been refined in~\cite{wang2019maximum}, which does include an explicit calculation of the contribution of the rim to the droplet surface energy. 

However, it is also of interest not only to understand the maximum spreading radius, but also,  the fluid dynamics during the spreading.  Hence, it is useful to examine the different stages of droplet impact. Stage one is where the droplet contacts the surface and no significant lamella is seen usually.  The lamella is formed (ejected) in the second stage and spreads with our without a rim being visible from initial moments.    The second stage ends when the droplet reaches its maximum radius.  The existence of the third stage, recoil of the lamella and formation of a sessile droplet, depends on the surface wettability~\cite{yarin2006drop}.

Modelling  stage two via a rim-lamella model provides another way of estimating $\maxR$, and is complementary to the energy-budget analysis.
Such models use a set of ordinary differential equations to describe the mass and momentum balances in the rim and lamella and hence, describe the evolution of the droplet spreading.  The first such model was introduced in~\cite{roisman2002normal} for an inviscid fluid; the model was later extended to the viscous fluids in~\cite{eggers2010drop}.
%


The rim-lamella model describes stage two; the model therefore requires initial conditions corresponding to this stage.
Different authors have use different approaches to formulating the initial conditions.  
Roisman et al. have modelled the ejection of the lamella using a force balance~\cite{roisman2002normal}.  Federchenko and Wang have used an energy-budget analysis~\cite{fedorchenko2004some}, while Eggers et al. have assigned initial conditions based on a geometric argument, and hence, without ascribing any $\mathrm{We}$- or $\mathrm{Re}$-dependence to the initial conditions~\cite{eggers2010drop}.  In the present work, we carefully review these assumptions and categorize them according to whether they are physically realistic, or merely expedient.


A recent set of theoretical~\cite{gordillo2019theory,garcia2020inclined} and experimental~\cite{Riboux2014} works has also sought to address the initial conditions for the rim-lamella formation.  The authors have performed a careful scaling analysis based on a force balance and derived an expression for the ejection time of the lamella from the initial droplet impact, in terms of $\myWe$ and $\myRe$.
    In these works, partial differential equations with a moving boundary were used to describe the mass and momentum fluxes in the lamella, which were then coupled to a system of ordinary differential equations for the fluid motion in the rim.  The equations can be analyzed an an expression for $\maxR$ derived whose functional form is very similar to that derived in~\cite{wildeman2016spreading},  using an energy-budget analysis.    This is a highly detailed theoretical approach, which is in close agreement with experimental results.  
		
A rim-lamella model is still a fundamental part of this line of work, and the proposed theoretical analysis is therefore complementary.
Furthermore, we include here for the first time a detailed proof showing why the standard rim-lamella models manifestly conserve mass.  This is a useful consistency check which must be passed by all models involving droplet impact and spreading.	
Finally, by applying the theory of estimates and perturbation theory to the rim-lamella model, the present paper builds on the previous theoretical analysis of the rim-lamella model carried out in~\cite{fedorchenko2004some}.   

\subsection{Plan of the Paper}

%
The paper is organized as follows.
In Section~\ref{sec:theory} we review the rim-lamella model in the  inviscid case and prove the self-consistency of the model under a stated range of initial conditions.  In Sections~\ref{sec:theory}--\ref{sec:perturbation} -- again under a stated range of initial conditions -- we place upper and lower bounds for the maximum spreading radius.  The derivations rely on  differential inequalities and perturbation theory and as such, are rigorous.  The upper and lower bounds are proportional to $\myWe^{1/2}$ in the limit of large Weber number.  Hence, by a sandwich result, the maximum spreading radius itself must scale as $\myWe^{1/2}$ in the same limit.  

The analysis of the rim-lamella model relies only on a minimal set of physically realistic assumptions.  However, there is some uncertainty concerning the choice of the initial conditions which feed into the rim-lamella model.  Some choices are made based on physical principles, others are made based on a combination of expediency and tradition.  To test the validity of our choices, we compare our predictions for the maximum spreading radius with experimental data and numerical simulations in Section~\ref{sec:experiments}.  Concluding remarks are given in Section~\ref{sec:conclusions}.

\section{Theoretical Analysis of the Rim-Lamella Model}
\label{sec:theory}

In this section we analyse the rim-lamella model introduced in~\cite{roisman2002normal}.  This model is concerned with the idealised inviscid limit where the boundary layer in the lamella plays no role in the rim evolution.  This is an important limiting case, and is the focus of the paper.  We first of all summarize the model before presenting our results.

\subsection{Summary of the Model}

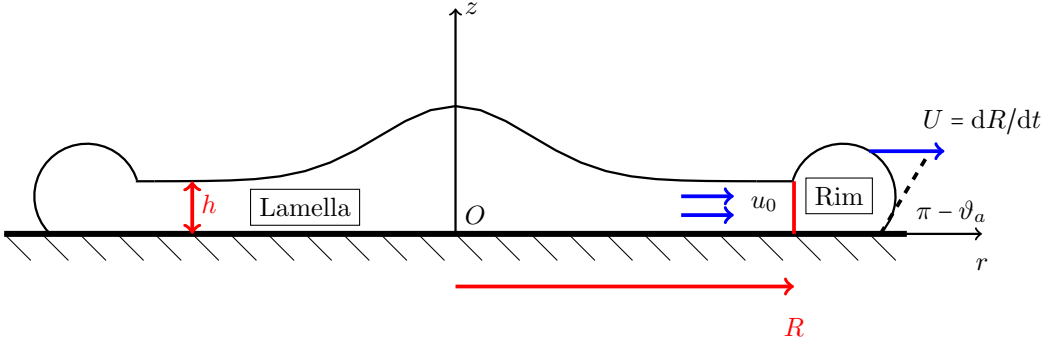
\begin{figure}[htb]
\begin{tikzpicture}
\draw[-,black,line width=0.8mm] (0,0) -- (12,0);
\foreach \x in {-1,...,22}
\draw (0.5+0.5*\x,0) -- (0.5+0.5*\x+0.5*0.707,-0.5*0.707);
\draw[->,black,line width=0.3mm] (6,0) -- (13,0);
\draw (13, -0.2) node[below] {$r$};
\draw[->,black,line width=0.3mm] (6,0) -- (6,3);
\draw (6, 3) node[right] {$z$};
\draw (6 ,0.25) node[right] {$O$};
%
%
\draw[-,black,line width=0.3mm] (1.75,0.7) -- (2,0.7);
\draw[color=black,line width=0.3mm,domain=2:10] plot (\normaltwo) node[right] {};
\draw[-,black,line width=0.3mm] (10,0.7) -- (10.5,0.7);
\draw [black,line width=0.3mm,domain=-45:162] plot ({11.15+0.7*cos(\x)}, {0.5+0.7*sin(\x)});
\draw [black,line width=0.3mm,domain=18:228] plot ({1.1+0.7*cos(\x)}, {0.5+0.7*sin(\x)});
\draw [<->,red,line width=0.5mm] (2.5,0) -- (2.5,0.7);
\draw (2.5, 0.4) node[right] {${\color{red}{h}}$}; 
\draw [->,red,line width=0.5mm] (6,-0.7) -- (10.5,-0.7);
\draw (10.5,-1) node[below] {${\color{red}{R}}$}; 
\draw [-,red,line width=0.5mm] (10.5,0) -- (10.5,0.7);
\draw (4, 0.6) node[rectangle,draw,below] {Lamella};
\draw (11.1, 0.75) node[rectangle,draw,below] {Rim};
\draw [black,dashed,line width=0.5mm] (11.65,0) -- (12.25,1);
\draw (12, 0.3) node[right] {$\pi-\thetadyn$};
\draw [->,blue,line width=0.5mm] (11.5,1.1) -- (12.5,1.1);
\draw (13,1.2) node[above] {$U=\mathd R/\mathd t$};
\draw [->,blue,line width=0.5mm] (7.5+1.5,0.25) -- (8.2+1.5,0.25);
\draw [->,blue,line width=0.5mm] (7.5+1.5,0.5) -- (8.2+1.5,0.5);
\draw (8.3+1.5,0.4) node[right] {$u_0$};
\end{tikzpicture}
\caption{Schematic diagram showing the cross-section of an axisymmetric rim-lamella structure}
\label{fig:schematic}
\end{figure}

Referring to Figure~\ref{fig:schematic}, and below the splash threshold, the droplet in the second stage of the motion is an axisymmetric structure made up of two regions, the lamella and the rim.  
In the lamella, the flow satisfies the following momentum-balance and mass-balance equations for the  depth-averaged flow $u(r,t)$ and height $h(r,t)$ of the lamella~\cite{yarin1995impact}:
\begin{subequations}
\begin{eqnarray}
\frac{\partial u}{\partial t}+u\frac{\partial u}{\partial r}&=&0,\label{eq:dudt}\\
\frac{\partial}{\partial t}\left(rh\right)+\frac{\partial}{\partial r}(u r h)&=&0\label{eq:dhdt},
\end{eqnarray}
\end{subequations}
valid for $t>\tau$, and $r\in (0,R)$. 
%
%
%
Here, $\tau$ is the start-time of the second stage of the motion and $R$ (depending on time) is the radius of the inner edge of the lamella.  The velocity equation~\eqref{eq:dudt} has the solution
\begin{equation}
u(r,t)=\frac{r}{t+t_0}.
\label{eq:urt}
\end{equation}
Here, $t_0$ is a parameter which is inherited from stage one of the motion, and is related to the time required for the pressure in the droplet to decay to zero~\cite{eggers2010drop}.
We further use the notation $u_0=R/(t+t_0)$ to denote the velocity at the inner edge of the lamella.
Equation~\eqref{eq:dhdt} for $h$ is less straightforward: any combination $h\propto (t+t_0)^{-2}f(r/(t+t_0))$ is a solution.  Roisman et al. first introduced the `engineering approximation for the drop height'~\cite{roisman2009inertia}:
\begin{equation}
\frac{h(r,t)}{R_0}=\frac{\eta}{(t+t_0)^2}\frac{R_0^2}{U_0^2}\mathe^{-(3\eta/4U_0^2) [r/(t+t_0)]^2}.
\label{eq:ht}
\end{equation}
which fits experimental data and numerical simulations well ($\eta$ is a dimensionless pre-factor).   We use Equation~\eqref{eq:ht} in the present work.  The factor of $3/4$ in the exponent ensures conservation of mass, which we describe in more detail below.

The flow in the lamella region drives the rim evolution; mass and momentum balance in the rim are described by the following ordinary differential equations, valid in the inviscid limit:
\begin{subequations}
\begin{eqnarray}
\frac{\mathd V}{\mathd t}&=& 2\pi R \left(u_0-U\right)h(R,t),\label{eq:mass}\\
V\frac{\mathd U}{\mathd t}&=&2\pi R \left[\left(u_0-U\right)^2 h(R,t) - \frac{\sigma}{\rho}\left(1-\cos\thetadyn\right)\right],\label{eq:mom}\\
\frac{\mathd R}{\mathd t}&=&U.
\end{eqnarray}%
\label{eq:roisman}%
\end{subequations}%
Here, $V$ is the rim volume, $U$ is the rim velocity, and $\sigma$ is the surface tension.    For  $\mathd R/\mathd t>0$, $\thetadyn$ is the advancing contact angle.  We  present results in the case of constant advancing contact angle, a reasonable assumption in a highly dynamic phase of spreading~\cite{yokoi2009numerical}.
Equation~\eqref{eq:roisman} is seeded with initial conditions at $t=\tau$:
\begin{equation}
R(\tau)=R_{init},\qquad U(\tau)=U_{init},\qquad V(\tau)=V_{init}.
\end{equation}
%
%
%
In the following analysis, it will be helpful to work with the velocity defect, $\Delta$:
\begin{equation}
\Delta=u_0-U=\frac{R}{t+t_0}-U.
\end{equation}
The momentum equation~\eqref{eq:mom} can then be re-written in terms of $\Delta$:
\begin{equation}
\frac{\mathd \Delta}{\mathd t}+\frac{\Delta}{t+t_0}=-\frac{2\pi R h}{V} \big\{\Delta^2-[c(t)]^2\big\}.
\label{eq:Deltadef}
\end{equation}
Here, we have introduced the characteristic speed
\begin{equation}
c(t)=\sqrt{\frac{\sigma(1-\cos\thetadyn)}{\rho h(R,t)}}=\sqrt{\frac{\sigma(1-\cos\thetadyn)}{\rho\eta R_0}}(U_0/R_0)(t+t_0)\mathe^{(3\eta/8U_0^2)[R/(t+t_0)]^2}.
\end{equation}
The characteristic speed is reminiscent of the Taylor--Culick speed for the retraction of a liquid sheet of thickness $h$, $c=\sqrt{2\sigma/\rho h}$.  A similar expression applicable to an expanding sheet over a solid surface can be derived by considering contact angle and balancing inertia and surface tension as in~\cite{roisman2002normal}.

\vspace{0.1in}
\noindent {\textbf{Remark:}} 
Equation~\eqref{eq:mom} with $V(\mathd{U}/\mathd t)$ on the left-hand side is one particular form of Newton's equation, $\text{mass}\times\text{acceleration}=\text{Force}$.  This form of Newton's equation is used in~\cite{roisman2002normal,eggers2010drop}.  In contrast, \cite{fedorchenko2004some}  uses a form of Newton's equation with $\mathd (UV)/\mathd t$  on the left-hand side.  However, by comparing the respective right-hand sides of these equations, it can be seen that they are identical.

\subsection{Restrictions on Initial Conditions and Key Result}

Physically, the rim-lamella model corresponds to a phase of the motion where the rim advances, but decelerates~\cite{Riboux2014}.  This places restrictions on the initial conditions of the rim-lamella model.  These are minimal assumptions required for physical realism.    As such, we assume that:
\begin{equation}
U_{init}>0,
\label{eq:boundIC0}
\end{equation}
such that the rim is advancing at $t=\tau$.  We also assume that
$0< \Delta(\tau)\leq c(\tau)$, hence:
\begin{equation}
0\leq \frac{R_{init}}{\tau+t_0}-U_{init}\leq c(\tau),
\label{eq:boundIC}
\end{equation}
such that $(\mathd U/\mathd t)_{t=\tau}\leq 0$.  Hence, the rim is decelerating at $t=\tau$.  Another physical interpretation is that the  velocity of the lamella relative to the rim  at $t=\tau$ should not exceed the characteristic velocity $c(\tau)$.  
%
%
%
A third restriction on the initial conditions is required to maintain the deceleration of the rim-lamella at later times $t\geq \tau$:
%
%
\begin{equation}
\frac{3\eta}{2U_0^2}\left[ \frac{R_{init}}{\tau+t_0}+\Delta(\tau)\right]^2<1.
\label{eq:boundIC2}
\end{equation}
This restriction is by no means obvious: it ensures that if at some time $c$ is an increasing function of time, the rate of increase is not too large.  This is made precise in Appendix~\ref{sec:AppA}. 

Using Gronwall's Inequality,   and the initial conditions~\eqref{eq:boundIC0}--\eqref{eq:boundIC2}, it can  be shown that  $\Delta(t)$ satisfies:
\begin{equation}
0< \Delta(\tau)\left(\frac{\tau+t_0}{t+t_0}\right)           \leq       \Delta \leq c,
\label{eq:boundDelta}
\end{equation}
Equation~\eqref{eq:boundDelta} is one of the main results of the present work.  It  ensures that $\Delta(t)\geq 0$ for all time up to the maximum spreading time $t_{max}$ when $U(t_{max})=0$.  It further ensures that $\Delta(t)\leq c(t)$.   Referring back to Equation~\eqref{eq:roisman}, 
this key result further ensures that deceleration of the rim is maintained at later times $t\geq 0$, as $V(\mathd U/\mathd t)=2\pi R\{\Delta^2-[c(t)]^2\}$.

The proof of Equation~\eqref{eq:boundDelta} relies on elementary methods, but is lengthy.  Further details are provided in Appendix~\ref{sec:AppA}.
The result in Equation~\eqref{eq:boundDelta} further ensures the self-consistency of the rim-lamella model, in the sense that the height of the rim exceeds the height of the lamella.  This self-consinstency is assumed in the earlier works such as~\cite{roisman2002normal} but is not shown explicitly.   We demonstrate this below in Section~\ref{sec:consistent}.

\subsection{Mass conservation in the Rim-Lamella Model}

We show that the total volume, $V_{tot}=V+V_{lamella}$, is a constant.  This is necessary for physical reaslism and it is assumed in various works on rim-lamella models~\cite{roisman2009inertia,eggers2010drop}.  However, it is useful to set out the steps in the proof, as doing this serves as justification for the various pre-factors in the engineering approximation~\eqref{eq:ht} for the drop height.  The starting-point for the calculation is the following expresison for the lamella volume:  
\[
V_{lamella}=2\pi \int_0^R h(r,t)r\,\mathd r.
\]
We use the expression~\eqref{eq:ht} for $h(r,t)$, to get:
\[
V_{lamella}=V_0 \left[1-\mathe^{-(3\eta/4U_0^2) [R/(t+t_0)]^2}\right].
\]
Here, $V_0=(4\pi/3)R_0^3$ is the initial volume of the droplet, prior to impact.  We compute:
\begin{eqnarray*}
\frac{\mathd V_{tot}}{\mathd t}&=&\frac{\mathd V}{\mathd t}+\frac{\mathd V_{lamella}}{\mathd t},\\
 &=&2\pi R(u_0-U)h(R,t)+
\left(\tfrac{4\pi}{3}R_0^3\right)\left(\frac{3\eta}{4 U_0^2}\right) \frac{2R}{(t+t_0)^2}\left(\frac{\mathd R}{\mathd t}-\frac{R}{t+t_0}\right)\mathe^{-(3\eta/4U_0^2) [R/(t+t_0)]^2},\\
&=&2\pi R(u_0-U)h(R,t)+2\pi R \left(\frac{\mathd R}{\mathd t}-\frac{R}{t+t_0}\right)h(R,t),\\
&=&0.
\end{eqnarray*}

\vspace{0.1in}
\noindent {\textbf{Remark:}} 
In the foregoing calculations, the factor of $3\eta/(4U_0^2)$ in the exponent in the engineering approximation for the drop height produces the exact cancellations such that $\mathd V_{tot}/\mathd t=0$. 

\subsection{Generic bounds for the Rim-Lamella model}

In this final sub-section, we draw together the arguments from the previous subsections, to produce a lower bound on $\max[R(t)]$.  The combination of this lower bound with a subsequent upper bound on $\max [R(t)]$ will then be used to prove rigorously the scaling behaviour $\max [R(t)]\sim \myWe^{1/2}$, via a sandwich result.
The starting-point is to look at $\Delta=R/(t+t_0)-\mathd R/\mathd t$.  By simple re-arrangement of this expression, we obtain:
\begin{equation}
\frac{\mathd R}{\mathd t}-\frac{R}{t+t_0}=-\Delta.
\label{eq:Rineq1}
\end{equation}
We apply the key bound~\eqref{eq:boundDelta}
\begin{equation}
\frac{\mathd R}{\mathd t}-\frac{R}{t+t_0}\leq  -\Delta(\tau)\frac{\tau+t_0}{t+t_0}.
\end{equation}
An application of Gronwall's inequality gives:
\begin{equation}
R(t)\leq \left[\frac{R_{init}}{\tau+t_0}-\Delta(\tau)\right](t+t_0)+\Delta(\tau)(\tau+t_0).
\end{equation}
Hence,
\begin{equation}
\frac{R(t)}{t+t_0}\leq \frac{R_{init}}{\tau+t_0}+\Delta(\tau).
\label{eq:Rtfin}
\end{equation}
We look again at Equation~\eqref{eq:Rineq1}.
Given the key inequalities~\eqref{eq:boundDelta} and~\eqref{eq:Rtfin},  this equation can be re-cast as an inequality:
\begin{equation}
\frac{\mathd R}{\mathd t}-\frac{R}{t+t_0} \geq -c(t)\geq
 -\left[\sqrt{ \frac{\sigma (1-\cos\thetadyn)}{\rho R_0 \eta} }(U_0/R_0)(\tau+t_0)\mathe^{(3\eta/8U_0^2)[R_{init}/(\tau+t_0)+\Delta(\tau)]^2} \right]\frac{t+t_0}{\tau+t_0}.
\label{eq:Rineq2}
\end{equation}
We call the term in square brackets  $\widehat{c}(\tau)$.  Applying Gronwall's inequality to Equation~\eqref{eq:Rineq2} then gives:
%
%
\begin{equation}
R(t) \geq \left(\frac{R_{init}}{\tau+t_0}\right)(t+t_0)-\frac{\widehat{c}(\tau)}{\tau+t_0}(t-\tau)(t+t_0).
\label{eq:Rtgt}
\end{equation}
Or,
\begin{equation}
R(t)\geq \left(\frac{R_{init}}{\tau+t_0}+\widehat{c}(\tau)\frac{\tau}{\tau+t_0}\right)t_0+\left(\frac{R_{init}}{\tau+t_0}+\widehat{c}(\tau)\frac{\tau-t_0}{\tau+t_0}\right)t-\frac{\widehat{c}(\tau)}{\tau+t_0}t^2.
\label{eq:RHS_Rstar}
\end{equation}

\vspace{0.1in}
\noindent {\textbf{Result:}} By taking the maximum of the RHS of Equation~\eqref{eq:RHS_Rstar} and expanding out quadratics, one obtains:
\begin{equation}
\max[R(t)]\geq R_*,\qquad R_*=\frac{\tau+t_0}{4\widehat{c}(\tau)}\left[\widehat{c}(\tau)+\frac{R_{init}}{\tau+t_0}\right]^2.\myqed
\label{eq:Rstar}
\end{equation}
%
Equation~\eqref{eq:Rstar} is a key equation which we use later to obtain an explicit theoretical bound on the maximum spreading radius, in terms of powers of $\myWe$.

\section{Self-Consistency for the Rim-Lamella Model}
\label{sec:consistent}

For the rim-lamella model to be consistent, it must be possible to construct a circular segment passing through the lamella extremity, and cutting the solid surface at the contact angle $\thetadyn$.  Furthermore, the maximum height attained by the circular segment must necessarily exceed the lamella height.  The idea is shown in Figure~\ref{fig:schematic}.  Only when this second condition is satisfied can the ODE model~\eqref{eq:roisman} be said to produce a genuine rim, and hence, to be physically self-consistent.

To prove that the rim-lamella model~\eqref{eq:roisman} does indeed exhibit this self-consistency, we combine an analysis of the ODE system with geometric arguments.  Our starting-point is to introduce the ratio:
\[
{\varphi}=\frac{V}{2\pi^2 h^2 R}.
\]
Let $A$ denote the cross-sectional area of the rim.  In the limit of large spreading radius $R$ greatly exceeds the thickness of the rim, the volume of the rim is well approximated by 
\begin{equation}
V=2\pi R A.
\label{eq:vapprox}
\end{equation}
Hence, $\varphi$ can be re-written as $\varphi= A/(\pi h^2)$.  

We derive an equation for $\varphi$ by direct differentiation of $\varphi=V/(2\pi^2 h^2 R)$.  We have:
\[
\frac{\mathd }{\mathd t}\left(\frac{V}{h^2 R}\right)=\frac{\mathd V}{\mathd t}\frac{1}{h^2 R}-\frac{2V}{Rh^3}\frac{\mathd h}{\mathd t}-\frac{V}{h^2 R^2}\frac{\mathd R}{\mathd t},\\
\]
By direct differentiation we obtain:
\[
\frac{\mathd }{\mathd t}\left(\frac{V}{h^2 R}\right)=\frac{2\pi\Delta}{h}+\left(\frac{V}{h^2 R}\right)(3u_0/R)+\left(\frac{V}{h^2 R}\right) (\Delta/R)\left[1-\frac{2\times 3\eta}{4U_0^2}\left(\frac{R}{t+t_0}\right)^2\right].
\]
Or,
\[
\frac{\mathd \varphi}{\mathd t}=\frac{\Delta}{\pi h}+\varphi(3u_0/R)+\varphi (\Delta/R)\left[1-\frac{3\eta}{2U_0^2}\left(\frac{R}{t+t_0}\right)^2\right].
\]
We use the bounds~\eqref{eq:boundIC2} and~\eqref{eq:Rtgt} to re-cast this as an inequality:
\[
\frac{\mathd\varphi }{\mathd t}\geq \frac{\Delta}{\pi h}+\varphi(3u_0/R)+\varphi (\Delta/R)\underbrace{\left[1-\frac{3\eta}{2U_0^2}\left(\frac{R_{init}}{\tau+t_0}+\Delta(\tau)\right)^2\right]}_{\geq 0, \text{ by inequalities~\eqref{eq:boundIC2} and~\eqref{eq:Rtgt}}}.
\]
We make further use of bounds (specifically, Equation~\eqref{eq:ht} for the drop height, and the inequality~\eqref{eq:boundDelta} for $\Delta$) to re-cast this as:
\[
\frac{\mathd \varphi}{\mathd t}\geq \frac{\Delta(\tau)}{\pi h_{init}}\left(\frac{t+t_0}{\tau+t_0}\right)+\frac{3}{t+t_0}\varphi.
\]
As this is a linear first-order differential inequality, we use Gronwall's inequality to get:
\[
\varphi(t)\geq \varphi(\tau)\left(\frac{t+t_0}{\tau+t_0}\right)^3+\frac{ \Delta(\tau)}{\pi h_{init}}\left(\frac{t+t_0}{\tau+t_0}\right)^2(t+t_0)\left[1-\frac{\tau+t_0}{t+t_0}\right].
\]
Thus, $\varphi(t)$ grows at least as fast as $t^3$; $\varphi(t)$ will exceed one even if $\varphi(\tau)=0$.  Put otherwise, there exists a time $t_*\geq \tau$ such that $\varphi(t)\geq 1$ for all $t>t_*$.  

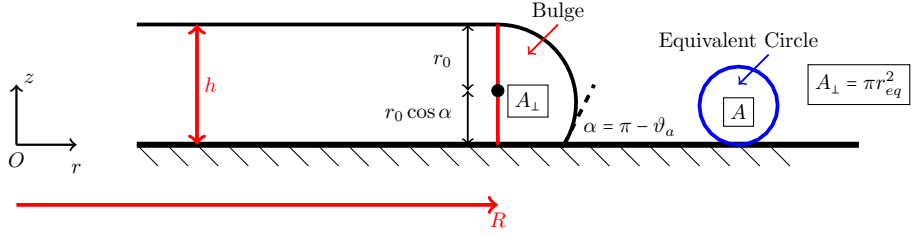
\begin{figure}
\centering
\begin{tikzpicture}[scale=0.8, transform shape]
%
%
%
\draw[-,black,line width=0.8mm] (0,0) -- (12,0);
\foreach \x in {-1,...,20}
\draw (0.5+0.5*\x,0) -- (0.5+0.5*\x+0.5*0.707,-0.5*0.707);

\draw[-,black,line width=0.5mm] (0,2) -- (6,2);

\draw [black,line width=0.5mm,domain=-35:90] plot ({6+1.3*cos(\x)}, {0.7+1.3*sin(\x)});
\draw (7, 2.5) node[below] {Bulge};
\draw [->,red,line width=0.25mm] (7,2) -- (6.5,1.5);
\draw (6.5, 1) node[rectangle,draw,below] {$A_\perp$};

\draw (10, 2) node[below] {Equivalent Circle};
\draw (12, 1) node[rectangle,draw] {$A_\perp=\pi r_{eq}^2$};
\draw [->,blue,line width=0.25mm] (10.5,1.5) -- (10,1);
\draw [blue,line width=0.5mm,domain=0:360] plot ({10+0.65*cos(\x)}, {0.65+0.65*sin(\x)});
\draw (10, 0.3) node[rectangle,draw,above] {$A$};
\draw [red,line width=0.5mm] (6,0) -- (6,2);
\draw (6, 0.9) node[circle,draw,fill=black,minimum size = 0.2cm, inner sep=0pt] {}; 
\draw [<->,black,line width=0.25mm] (5.5,0.9) -- (5.5,2);
\draw (4.8, 1.4) node[right] {$r_0$};
\draw [<->,black,line width=0.25mm] (5.5,0) -- (5.5,0.9);
\draw (4, 0.5) node[right] {$r_0\cos\alpha$};
\draw [black,dashed,line width=0.5mm] (7.1,0) -- (7.6,1);
\draw (7.3, 0.3) node[right] {$\alpha=\pi-\thetadyn$};
\draw [<->,red,line width=0.5mm] (1,0) -- (1,2);
\draw (1, 1) node[right] {${\color{red}{h}}$}; 
\draw[->,black,line width=0.3mm] (-2,0) -- (-1,0);
\draw (-1, -0.2) node[below] {$r$};
\draw[->,black,line width=0.3mm] (-2,0) -- (-2,1);
\draw (-2, 1.1) node[right] {$z$};
\draw (-2 ,0) node[below] {$O$};
\draw [->,red,line width=0.5mm] (-2,-1) -- (6,-1);
\draw (6, -1) node[below] {${\color{red}{R}}$}; 
\end{tikzpicture}
\caption{Schematic diagram showing the putative `no-rim' configuration}
\label{fig:norim}
\end{figure}
We now show that the rim-lamella model is self-consistent.   Suppose instead for a contradiction that a situation such as the one in Figure~\ref{fig:norim} pertains for all time $t>\tau$.   Here, the cross-sectional area $A_\perp$ is a circular segment which connects smoothly to the rim (no cusp point), and  makes an angle $\thetadyn$ with the solid surface.  We present the argument for the case $\thetadyn>\pi/2$, the case $\thetadyn\leq \pi/2$ is similar.  
We look at the cross-sectional area $A_\perp$ and construct an equivalent circle with the same area $A_\perp$ and radius $r_{eq}$ (hence, $A_\perp=\pi r_{eq}^2$).   Hence,
\begin{equation}
\frac{r_{eq}^2}{h^2}=\frac{\thetadyn-\sin\thetadyn\cos\thetadyn}{2\pi(1-\cos\thetadyn)^2},\qquad \thetadyn>\pi/2.
\end{equation}
It can be checked that $r_{eq}^2/h^2<1$ for all values of $\thetadyn$.  Thus,
\begin{equation}
\varphi=\frac{V}{2\pi^2 R h^2}=\frac{ A_{\perp}}{\pi h^2}=\frac{r_{eq}^2}{h^2}<1,
\end{equation}
for all time $t\geq \tau$.
However, $\varphi(t)\geq 1$ for $t>t_*$.  This is a contradiction.  Therefore, the situation in Figure~\ref{fig:norim} cannot pertain and hence, the rim-lamella model is self-consistent.
Furthermore, the scenario in Figure~\ref{fig:norim} is a limiting case.  In any other similar scenario where the bulge failed to produce a rim, the bulge area $A$ would satisfy $A<A_\perp$ (the rim and the lamella would then join in a cusp point).  This would again generate the contradiction $\varphi<1$ for all time $t\geq \tau$.

\section{Theoretical Lower Bound on the Spreading Radius}
\label{sec:bounds}

In this section we obtain a theoretical lower bound for the maximum spreading radius.  To obtain concrete results, instead of working with the engineering approximation for the drop height (Equation~\eqref{eq:ht}), we use the `remote asymptotic solution'
\begin{equation}
\frac{h(R,t)}{R_0}=\frac{\eta}{(t+t_0)^2}\frac{R_0^2}{U_0^2}=\frac{h_{init}}{R_0}\left(\frac{\tau+t_0}{t+t_0}\right)^2.
\label{eq:RAS}
\end{equation}
In the studies in the literature, little or no difference is seen in the model solutions which use either the engineering approximation or the remote asymptotic solution.  We emphasize here that rigorous bounds on the spreading radius are possible using either approach, but the bounds are particularly analytically tractable in the case of the remote asymptotic solution.  
The lower bound for $\max [R(t)]$ derived earlier in case of the engineering approximation for the drop height still stands, if the remote asymptotic solution is used instead.  Indeed, the lower bound 
for $\max [R(t)]$ now takes a simpler form, which we state as follows.
%
%

\noindent {\textbf{Result:}}  
The rim-lamella model~\eqref{eq:roisman} with $h(R,t)$ given by the remote asymptotic solution admits the following inequalities:
\begin{subequations}
\begin{eqnarray}
0&\leq & \Delta(\tau)\left(\frac{\tau+t_0}{t+t_0}\right)           \leq       \Delta \leq c(t),\\
R(t)&\leq &\left[\frac{R_{init}}{\tau+t_0}-\Delta(\tau)\right](t+t_0)+\Delta(\tau)(\tau+t_0),\\
\max[R(t)]&\geq & R_*,
\end{eqnarray}
\end{subequations}
where
\begin{equation}
R_*=\frac{\tau+t_0}{4{c}(\tau)}\left[{c}(\tau)+\frac{R_{init}}{\tau+t_0}\right]^2.
\label{eq:Rstar_RAS}
\end{equation}
Furthermore, the rim-lamella model remains self-consistent, in the sense of Section~\ref{sec:consistent}.\myqed
%
%

We extract the explicit scaling behaviour implied by this result using assumptions on the initial conditions for the rim-lamella model.  At sufficiently high Weber number, it can be argued~\cite{fedorchenko2004some} that $R_{init}$ and $\tau+t_0$ are Weber-number independent, as the development of the lamella prior to rim generation is due to inertial effects only.  This yields $R_*\geq (1/4)R_{init}^2/[c(\tau)(\tau+t_0)]$, thereby establishing the scaling behaviour:
\begin{equation}
\max [R(t)]\geq (\text{Const.})\times\myWe^{1/2}.
\label{eq:qualitative}
\end{equation}

\subsection{Developing a quantitative lower bound}

It is worthwhile not only to establish a qualitative result such as~\eqref{eq:qualitative}, but a quantiatitve one also, that can be used to compare against experiments and direct numerical simulations.
  In order to do this, it is necessary to have closed-form expressions for $R_{init}$ and $\tau+t_0$.    In contrast to the restrictions which are set out on the initial conditions in Equations~\eqref{eq:boundIC0}--\eqref{eq:boundIC2}, which are justified \textit{a priori}, the required assumptions on the initial conditions more expedient. However, they are justified by reference to the literature, and are consistent with 
Equations~\eqref{eq:boundIC0}--\eqref{eq:boundIC2}:
\begin{description}
\item[Assumption 1] As the rim volume is small compared to the lamella volume at time $t=\tau$, we use $V_{init}=0$.  This follows~\cite{roisman2002normal,fedorchenko2004some}.
\item[Assumption 2]  To avoid a singularity in the momentum equation at $t=\tau$, which would follow from $V_{init}=0$,  we use $\Delta(\tau)=c(\tau)$. 
This follows~\cite{roisman2002normal} and is implicit in~\cite{eggers2010drop}.
\item[Assumption 3] A consistency condition on the lamella vertical velocity at $r=0$, $(\partial h/\partial t)_{r=0}=-U_0$.  This follows~\cite{fedorchenko2004some}, and can be rewritten as:
\begin{equation}
\frac{2h_{init}}{\tau+t_0}=U_0,
\label{eq:eb1}
\end{equation}
\end{description}
Assumptions 1--2 are translated into initial conditions as follows:
\begin{equation}
V_{init}=0,\qquad \Delta(\tau)=c(\tau),\qquad U_{init}=\frac{R_{init}}{\tau+t_0}-c(\tau)>0.
\label{eq:RAS_ICs}
\end{equation}
%
%
%
Based on these assumptions, we perform an energy balance to connect $E_0$, the initial energy prior to impact, with the energy of the droplet at the onset of the second stage of the droplet spreading.  
To simplify the energy-budget calculation, we make a further assumption.
\begin{description}
\item[Assumption 4]  The initial droplet configuration at $t=\tau$ is that of a `pancake'.  
\end{description}
This amounts to applying the remote asymptotic solution across the entirety of the spreading droplet.
%
The initial kinetic energy is therefore determined by the flow in the lamella, with $u=r/(\tau+t_0)$ and $w=-2z/(\tau+t_0)$, at $t=\tau$. Hence,
\[
\text{K.E}=2\pi \int_0^{R_{init}} r\mathd r \int_0^{h_{init}} \tfrac{1}{2}\rho \left[ \left(\frac{r}{\tau+t_0}\right)^2+\left(\frac{2z}{\tau+t_0}\right)^2\right]\mathd z.
\]
We use $\pi R_{init}^2 h_{init}=V_0$ to re-write this as:
\[
\text{K.E.}=\rho V_0\left[\tfrac{1}{4}\left(\frac{R_{init}}{\tau+t_0}\right)^2+\tfrac{1}{6}\left(\frac{2 h_{init}}{\tau+t_0}\right)^2\right].
\]
The energy balance at $t=\tau$ therefore reads:
\begin{equation}
\rho V_0\left[\tfrac{1}{4}\left(\frac{R_{init}}{\tau+t_0}\right)^2+\tfrac{1}{6}\left(\frac{2 h_{init}}{\tau+t_0}\right)^2\right]+\sigma \pi R_{init}^2 +\frac{2\sigma V_0 }{ R_{init}}=E_0,
\label{eq:eb0}
\end{equation}
where $E_0=(1/2)\rho V_0U_0^2+4\pi \sigma R_0^2$ is the initial energy of the droplet, prior to impact.  
%

%
We combine Equations~\eqref{eq:eb0}--\eqref{eq:eb1}.  In non-dimensional terms, with $\widetilde{R}_{init}=R_{init}/R_0$, we obtain:
\begin{equation}
\tfrac{3^2}{2^8}\widetilde{R}_{init}^6+\tfrac{3}{4\myWe}\widetilde{R}_{init}^2+\tfrac{2}{\myWe}\frac{1}{\widetilde{R}_{init}}=\tfrac{1}{3}+\tfrac{3}{2\myWe}.
\label{eq:Rest0}
\end{equation}
From Equation~\eqref{eq:Rest0}, and for large $\myWe$, the initial droplet radius is  Weber-number independent, and reads:
\begin{equation}
R_{init}/R_0=\tfrac{2^{4/3}}{3^{1/2}}.
\label{eq:Rinit_est}
\end{equation}
a result  previously obtained in~\cite{fedorchenko2004formation}.  Also,
\begin{equation}
\frac{R_{init}}{\tau+t_0}=\tfrac{3}{8}U_0\left(R_{init}/R_0\right)^3=\tfrac{2}{\sqrt{3}}U_0.
\label{eq:u0init_est}
\end{equation}
Hence,
\begin{equation}
U_{init}/U_0= \left[\frac{R_{init}}{\tau+t_0}-c(\tau)\right]/U_0=
\tfrac{2}{\sqrt{3}}-2^{1/3}\myWe^{-1/2}(1-\cos\thetadyn)^{1/2},\qquad \myWe\gg 1.
\end{equation}
Thus, $U_{init}>0$ for $\myWe\gg 1$.
%
%
From Equation~\eqref{eq:Rstar_RAS}, we have:
%
\begin{multline*}
R_{max}\geq R_*= \frac{\tau+t_0}{4c(\tau)}\left[c(\tau)+\frac{R_{init}}{\tau+t_0}\right]^2\\
\geq  \tfrac{1}{2}R_{init}+\frac{R_{init}^2}{4c(\tau)(\tau+t_0)}=\tfrac{1}{2}R_{init}+\tfrac{1}{4}\frac{R_{init}}{\tau+t_0} \frac{\sqrt{\rho V_0}} {\sqrt{\pi\sigma(1-\cos\thetadyn)}}.
\end{multline*}
We use Equations~\eqref{eq:Rinit_est}--\eqref{eq:u0init_est} and conclude:
\begin{equation}
R_{max}/R_0 \geq  	 \tfrac{2^{1/3}}{3^{1/2}}+ \tfrac{1}{3}\myWe^{1/2}(1-\cos\thetadyn)^{-1/2}.
\label{eq:R_max_bound_XXX}
\end{equation}

\subsection{Validity of the Assumptions}

Assumption 1 ($V_{init}=0$) is consistent with experiments~\cite{qin2019viscosity}.  However, the other assumptions are not likely to be in exact agreement with experiments.  In particular:
\begin{itemize}[noitemsep]
\item The validity of Assumption 2 is not known from experiments, although it is consistent with the deceleration of the lamella.
\item Assumption 3 is simply a convenient approximation for the shape of the rim at the onset of the rim formation;
\item Assumption 4: As the droplet droplet dome falls and comes ever closer to the wall, the vertical speed gradually decreases from $U_0$ due to the influence of the wall~\cite{cheng2022drop}.  Hence, Equation~\eqref{eq:eb1} likely over-estimates $h_{init}$.
\end{itemize}
However, these assumptions are expected to give the correct order-of-magnitude estimates for $R_{init}$ and $\tau+t_0$.  This is tested below in Section~\ref{sec:experiments}, where we compare the predictions of the model with data available from experiments and simulations.

\subsection{Correction due to the rim}

Following a standard approach in the literature~\cite{eggers2010drop,roisman2002normal}, we approximate the cross-sectional area of the rim as a circular segment.  By standard geometric arguments, the cross-sectional area of the rim can therefore be computed as $A_\perp=(1/2)r_0^2\left[2\thetadyn-\sin(2\thetadyn)\right]$,
%
where $r_0$ is the radius of the circle.  Hence, the footprint of the rim can be estimated as $2a$, where $a=r_0\sin\thetadyn$.  Hence, 
\begin{equation}
a=\sin(\thetadyn)\sqrt{\frac{2A_\perp}{2\thetadyn-\sin(2\thetadyn)}}.
\end{equation}
Furthermore, the volume of the rim can be approximated as $V\approx 2\pi R A_{\perp}$ (\textit{cf}. Equation~\eqref{eq:vapprox}).  Finally, this gives an approximation for the droplet radius, $\mathcal{R}$, $\mathcal{R}\approx R+2a$.
Hence,
\begin{equation}
\mathcal{R}\approx R+2\sin\thetadyn\sqrt{  \frac{V}{\pi R [2\thetadyn-\sin(2\thetadyn)]}}.
\end{equation}

\section{Theoretical Upper Bound on the Spreading Radius}
\label{sec:perturbation}

Following the previous section (Section~\ref{sec:bounds}, Theoretical Lower Bound), in this section we develop a theoretical upper bound on the spreading radius.  The upper bound and the lower bound both exhibit the $\myWe^{1/2}$ scaling at large Weber number.  In this way, we develop a `sandwich result', such that the spreading radius exhibits the same scaling behaviour.
We  use Assumptions 1--2 from Section~\ref{sec:bounds}, together with the remote asymptotic assumption.  We do not use, or in any other way rely on Assumptions 3--4.
In this way,  the initial condition $V_{tot}(\tau)=V_{lamella}(\tau)=\pi R_{init}^2 h_{init}$ can be used to collapse rim-lamella model~\eqref{eq:roisman} into a single equation:
\begin{equation}
\label{eq:ODE_R_mdb}
\frac{\mathd}{\mathd t}\bigg\{(t+t_0)^2 \frac{\mathd}{\mathd t} \left[\frac{R}{t+t_0}-\frac{(\tau+t_0)^2}{3 R_{init}^2}\left(\frac{R}{t+t_0}\right)^3\right]\bigg\} = - \frac{2 c(\tau)^2}{R_{init}^2} (t+t_0)^2 \frac{R}{t+t_0},\qquad t>\tau.
\end{equation}
(we show the details of this calculation in Appendix~\ref{app:collapse}).  This equation is reminiscent of the single equation for the spreading radius derived in~\cite{villermaux2011drop} for droplet impact on a small solid sphere.  In that work, the equation for the spreading radius was reduced to that of simple harmonic motion.  We follow a similar approach here, to show that Equation~\eqref{eq:ODE_R_mdb} admits a solution in terms of a perturbation expansion, the lowest order of which corresponds to simple harmonic motion.

Equation~\eqref{eq:ODE_R_mdb} is close to a family of integrable systems found in Astrophysics (the Emden-Fowler Equation~\cite{horedt1986exact,zaitsev2002handbook}) and as such it can be solved by a simple perturbation method. To motivate the method, notice that the term in the square bracket in the LHS consists of a sum of two terms:
$$\frac{R}{t+t_0}-\frac{(\tau+t_0)^2}{3 R_{init}^2}\left(\frac{R}{t+t_0}\right)^3 = \frac{R}{t+t_0} \left(1 - \frac{V_{lamella}}{3 V_{tot}}\right).$$
The idea is that, although initially the ratio $V_{lamella}/(3 V_{tot})$ is equal to $1/3$, it rapidly decreases down to smaller values so one can assume this ratio is small and perform a series expansion by introducing a parameter $\mysmall$ (to be set  to 1 at the end) into Equation~\eqref{eq:ODE_R_mdb}, which now reads
\begin{equation}
\frac{\mathd}{\mathd t}\bigg\{(t+t_0)^2 \frac{\mathd}{\mathd t} \left[\frac{R}{t+t_0}-\mysmall \frac{(\tau+t_0)^2}{3 R_{init}^2}\left(\frac{R}{t+t_0}\right)^3\right]\bigg\} = - \frac{2 c(\tau)^2}{R_{init}^2}  (t+t_0)^2 \frac{R}{t+t_0}\,.
\end{equation}
The perturbative solution to this system is written in a few steps. First, define the new variable $L(t)$ in terms of precisely the quantity in square brackets discussed above: 
\begin{equation}
\label{eq:def_L_mdb}
\frac{L}{t+t_0} := \frac{R}{t+t_0}-\mysmall \frac{(\tau+t_0)^2}{3 R_{init}^2}\left(\frac{R}{t+t_0}\right)^3	
\,.
\end{equation}
In this way, Equation~\eqref{eq:ODE_R_mdb} becomes
\begin{equation}
\frac{\mathd}{\mathd t}\left[(t+t_0)^2 \frac{\mathd}{\mathd t} \left(\frac{L}{t+t_0}\right)\right] = - \frac{2 c(\tau)^2}{R_{init}^2} (t+t_0)^2 \frac{R}{t+t_0}\,,
\end{equation}
or, developing the derivatives and rearranging,
\begin{equation}
\label{eq:HO_mdb}
\frac{\mathd^2 L}{\mathd t^2} + \Omega^2 {R} = 0\,,\qquad \Omega := \sqrt{2} \,\frac{c(\tau)}{R_{init}}\,.
\end{equation}
It is remarkable how this exact equation looks like the harmonic oscillator, with fundamental frequency $\Omega$. However, one needs to find $R$ in terms of $L$ and that is the second step: we solve equation (\ref{eq:def_L_mdb}) for $R/(t+t_0)$. Formally, we are solving the ``Casus Irreducibilis'' of the cubic equation in the case when $L$ is positive but `small'. The solution gives the power series:  
\begin{equation}
\label{eq:def_R_mdb}
R = {L} \sum_{k=0}^\infty \left(\frac{\mysmall}{3}\right)^{k} B_k \left(\frac{\tau+t_0}{R_{init}}\right)^{2k} \left(\frac{L}{t+t_0}\right)^{2k} = L + \mysmall B_1 \frac{(\tau+t_0)^2}{3 R_{init}^2 (t+t_0)^2} L^3 + \ldots \,,
\end{equation}
where the coefficients $B_k$ are integers in a non-decreasing sequence, with limiting ratio $\lim_{k\to\infty} B_{k+1}/B_k = 27/4$. The first few are given by 
$$B_0 = B_1 = 1, \quad B_2 = 3, \quad B_3 = 12, \quad B_4 = 55, \quad B_5 = 273, \quad B_6 = 1428.$$
It can be shown that the series (\ref{eq:def_R_mdb}) converges when $V_{lamella}<V_{tot}$, so the method is justified by the physics of mass conservation.

In the third and last step of the method, we write a perturbative solution of the form
\begin{equation}
\label{eq:L_pert_mdb}
L(t) = \sum_{k=0}^\infty \mysmall^k L_k(t) = L_0(t) + \mysmall L_1(t) + \mysmall^2 L_2(t) + \ldots\,.
\end{equation}
Replacing this into the ODE system (\ref{eq:HO_mdb})--(\ref{eq:def_R_mdb}), we get a hierarchy of harmonic oscillator equations, one for each order in the expansion in powers of $\mysmall$:
\begin{eqnarray}
\label{eq:L_0_eq}
{\mathcal{O}}(\mysmall^0): &\qquad& \frac{\mathd^2 L_0}{\mathd t^2} + \Omega^2 L_0 = 0\,,\\
\label{eq:L_1_eq}
{\mathcal{O}}(\mysmall^1): &\qquad &\frac{\mathd^2 L_1}{\mathd t^2} + \Omega^2 L_1  = - \frac{(\tau+t_0)^2\Omega^2}{3 R_{init}^2 (t+t_0)^{2}} {(L_0)^3}{}\,,\\
\label{eq:L_2_eq}
{\mathcal{O}}(\mysmall^2): &\qquad& \frac{\mathd^2 L_2}{\mathd t^2} + \Omega^2 L_2  = \frac{(\tau+t_0)^2}{R_{init}^2(t+t_0)^{2}} (L_0)^2 \frac{\mathd^2 L_1}{\mathd t^2}\,,\\
\label{eq:L_3_eq}
{\mathcal{O}}(\mysmall^3): &\qquad& \frac{\mathd^2 L_3}{\mathd t^2} + \Omega^2 L_3  = - \frac{(\tau+t_0)^2}{R_{init}^2(t+t_0)^{2}} L_0 \left[L_0 \frac{\mathd^2 L_2}{\mathd t^2} - \Omega^{-2} \left(\frac{\mathd^2 L_1}{\mathd t^2}\right)^2\right]\,,
\end{eqnarray}
the next equations' RHSs becoming naturally more complex as the order increases but they are all hierarchical and solvable, being harmonic oscillators with known forcing functions. Notice that these forcing functions decay as $t$ grows, which on the one hand prevents the appearance of resonances (secular terms) and on the other hand makes the perturbative solution converge quickly.

The initial conditions for this system are determined from the initial conditions at $t = \tau$ for the rim radius $R$ and its time derivative. We have
$$R(\tau) = R_{init}, \qquad \frac{\mathd R}{\mathd t}(\tau) = \frac{R_{init}}{\tau+t_0} - c(\tau) 
\,.$$
Now, from equation (\ref{eq:def_L_mdb}), we can evaluate $L(\tau)$ and also $(\mathd L/\mathd t)_{t=\tau}$, giving 
\begin{eqnarray*}
L(\tau) &=& {R_{init}}-\frac{\mysmall}{3}{R}_{init}\,, \\
\frac{\mathd L}{\mathd t}(\tau) &=& \frac{R_{init}}{\tau+t_0} - c(\tau) - \frac{\mysmall}{3}\left[\frac{R_{init}}{\tau+t_0} -3\, c(\tau)\right]\,.
\end{eqnarray*}
Comparing this with the perturbative expansion (\ref{eq:L_pert_mdb}) we get the set of initial conditions
\begin{eqnarray}
\label{eq:L_0_IC}
L_0(\tau) = R_{init}, &\qquad &\frac{\mathd L_0}{\mathd t}(\tau)  = \frac{R_{init}}{\tau+t_0} - c(\tau)\,,\\
\label{eq:L_1_IC}
L_1(\tau) = -\frac{R_{init}}{3}, &\qquad& \frac{\mathd L_1}{\mathd t}(\tau)  = c(\tau) -\tfrac{1}{3}\left(\frac{R_{init}}{\tau+t_0}\right)\,,\\
\label{eq:L_k_IC}
L_k(\tau) = 0\,,  &\qquad& \frac{\mathd L_k}{\mathd t}(\tau) = 0\,, \qquad k \geq 2\,.
\end{eqnarray}

The solution to the lowest-order equation~\eqref{eq:L_0_eq} with the prescribed initial conditions~\eqref{eq:L_0_IC} is easily found:
\[
L_0(t) =R_{init} \cos(\Omega(t-\tau)) + \frac{R_{init}}{\sqrt{2}} \left[\frac{R_{init}}{(\tau+t_0) c(\tau)} - 1 \right] \sin(\Omega(t-\tau)) =: M_0 \cos(\Omega(t-\tau)+\phi_0),
\]
where 
\[
M_0 := R_{init} \sqrt{1+\tfrac{1}{2} \left[\frac{R_{init}}{(\tau+t_0) c(\tau)} - 1 \right]^2}\,,\qquad \phi_0 := - \arctan\bigg\{\tfrac{1}{\sqrt{2}}\left[\frac{R_{init}}{(\tau+t_0) c(\tau)} - 1\right]\bigg\},
\]
where we notice that $-\pi/2\leq \phi_0 < 0$ by virtue of condition (\ref{eq:RAS_ICs}) on the positivity of $U_{init}$.
With this solution, to lowest order in $\mysmall$ one can estimate the upper bound $R_{\max}$ to be equal to $M_0$ and the time $t_{\max}$ at which this maximum occurs:
\begin{equation}
R_{\max} \sim  R_{init} \sqrt{1+\tfrac{1}{2} \left[\frac{R_{init}}{(\tau+t_0) c(\tau)} - 1 \right]^2}\,,\qquad t_{\max} = \tau +\Omega^{-1} \arctan\bigg\{\tfrac{1}{\sqrt{2}}\left[\frac{R_{init}}{(\tau+t_0) c(\tau)} - 1\right]\bigg\}.
\label{eq:Rmax_upper0}
\end{equation}
In the large-Weber-number limit, $c(\tau)$ is small, and $R_{init}$ and $\tau+t_0$ are Weber-number independent (Section~\ref{sec:bounds}).  Hence, and these formula have the expected scaling:
\begin{equation}
R_{\max} \sim  \frac{R_{init}^2}{\sqrt{2}\, (\tau+t_0) c(\tau)} \sim {\mathrm{We}}^{1/2}\,, \qquad t_{\max} \sim \tau + \frac{\pi R_{init}}{2\sqrt{2} \,c(\tau)} \sim {\mathrm{We}}^{1/2}.
\label{eq:Rmax_upper}
\end{equation}
Recalling that this approach gives an upper bound for $\max [R(t)]$, we have:
\begin{equation}
\max [R(t)]\leq (\text{Const.})\times\myWe^{1/2}.
\label{eq:qualitative1}
\end{equation}
at sufficiently high Weber number, provided $V_{init}=0$.

\vspace{0.1in}
\noindent {\textbf{Remark:}} In the same limit,
\begin{equation}
\frac{R_{max}}{U_0 t_{max}}=\frac{2}{\pi}\frac{R_{init}}{(\tau+t_0)U_0}.
\label{eq:Rmax_upper1}
\end{equation}
Reading off from Equation~\eqref{eq:u0init_est}, this gives $R_{max}/(U_0 t_{max})=4/(\pi \sqrt{3})\approx 0.7351$.

The solution to the next order is found in a similar way. The method of variation of parameters gives the particular solution
\begin{multline}
L_{1,part}(t) = \Omega^{-1} \sin(\Omega (t-\tau)) \int_{\tau}^t F_1(t')\cos(\Omega(t'-\tau)) \mathd t' \\- \Omega^{-1}\cos(\Omega (t-\tau)) \int_{\tau}^t F_1(t') \sin(\Omega(t'-\tau))\mathd t',
\end{multline}
where $F_1$ is the RHS of Equation~\eqref{eq:L_1_eq}, so:
$$\Omega^{-1} F_1(t) = - {\sqrt{2}\,c(\tau)(\tau+t_0)^2}  \bigg\{{1+\tfrac{1}{2} \left[\frac{R_{init}}{(\tau+t_0) c(\tau)} - 1 \right]^2}\bigg\}^{\frac{3}{2}} \frac{\cos^3(\Omega(t-\tau)+\phi_0)}{3(t+t_0)^{2}}\,.$$
This particular solution is such that $L_{1,part}(\tau) = \left({\mathd L_{1,part}}/{\mathd t}\right)_{t=\tau} = 0$. Therefore, to implement the initial conditions (\ref{eq:L_1_IC}) we use the homogeneous solution, giving:
$$L_1(t) = L_{1,part}(t)  -  \tfrac{1}{3}{R}_{init} \cos(\Omega(t-\tau)) + \tfrac{1}{\sqrt{2}}R_{init} \bigg\{1-\tfrac{1}{3} \left[\frac{R_{init}}{(\tau+t_0)c(\tau)}\right]\bigg\} \sin(\Omega(t-\tau))\,.$$
Higher-order solutions are similarly found, with the extra simplifying result that only a particular solution needs to be found, due to the initial conditions (\ref{eq:L_k_IC}).
After these solutions for $L_0, L_1, \ldots$ are found, the solution to the original variable $R(t)$ is obtained simply by reverting back to Equation~\eqref{eq:HO_mdb}:
\begin{equation}
\label{eq:R_of_L_mdb}
R(t) = - \Omega^{-2} \frac{\mathd^2 L}{\mathd t^2} =  - \Omega^{-2} \left(\frac{\mathd^2 L_0}{\mathd t^2} + \mysmall \frac{\mathd^2 L_1}{\mathd t^2} +  \mysmall^2 \frac{\mathd^2 L_2}{\mathd t^2} + \mysmall^3 \frac{\mathd^2 L_3}{\mathd t^2} + \ldots\right)\,.
\end{equation}
In the next section we show a comparison between this analytical solution and the numerical solutions of the full rim-lamella model~\eqref{eq:roisman}.

\section{Comparison of Theoretical Analysis with Numerical Solutions; Comparison of Numerical Simulations with Experiments / DNS}
\label{sec:experiments}

In this section, we examine the sharpness of our bounds on the maximum spreading radius by comparing our results with numerical solutions of the full rim-lamella model~\eqref{eq:roisman}.  We also instigate the validity of our general approach by comparing our predictions for the maximum spreading radius with data from experiments and direct numerical simulation (DNS).
To generate quantitative results, we use the methodology in Section~\ref{sec:bounds} to produce quantitative estimates of the initial conditions on $R_{init}$ and $\tau+t_0$.

\subsection{Comparison between the Theoretical Analysis and Numerical Solutions of the Rim-Lamella Model}

\begin{figure}
	\centering
		\includegraphics[width=0.5\textwidth]{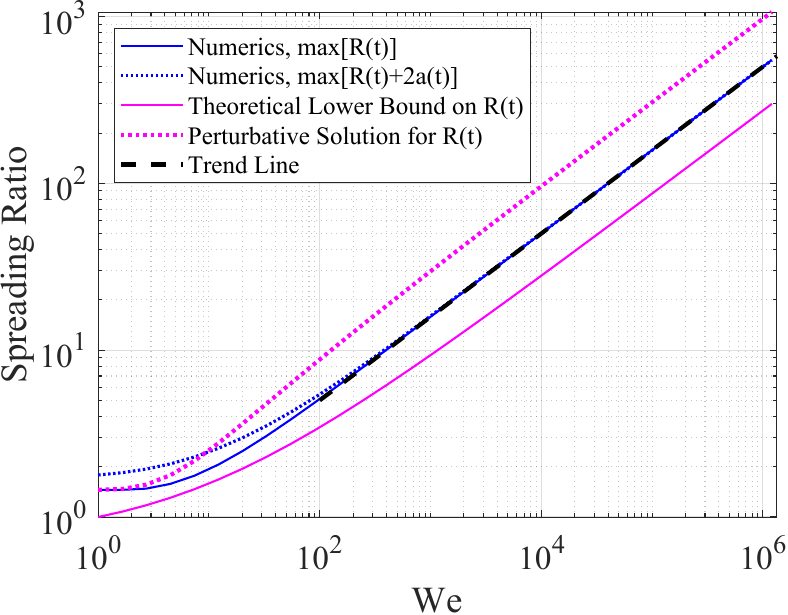}
		\caption{Estimates of the maximum spreading ratio as a function of $\myWe$: comparison of theoretical upper bound~\eqref{eq:R_max_bound_XXX} with numerical solutions ($\thetadyn=2\pi/3$).  The dashed line is a trend-line $\maxR/R_0=(\text{Const.})\times \mathrm{We}^{1/2}$; this has been added to guide the eye.}
	\label{fig:scaling1}
\end{figure}

We solve the rim-lamella model~\eqref{eq:roisman} numerically. We enumerate the initial conditions for the model, the methodology for deriving which has already been described in Section~\ref{sec:bounds}:
\begin{enumerate}[noitemsep]
\item We use the remote asymptotic solution~\eqref{eq:RAS} for $h(R,t)$.
\item We use $V_{init}=\epsilon$, where $\epsilon=10^{-4}$ or $10^{-6}$.  We have checked that the results do not depend on the choice of the (small) value of $\epsilon$. 
\item We use $R_{init}/R_0=2^{4/3}/3^{1/2}$, from Equation~\eqref{eq:Rinit_est}.  We also use $\pi h_{init}R_{init}^2=V_0-\epsilon$.
\item We further use initial conditions derived from Equation~\eqref{eq:RAS_ICs}, specifically:
\begin{equation}
\Delta(\tau)=c(\tau)\implies U_{init}=\frac{R_{init}}{\tau+t_0}-c(\tau).
\label{eq:ICs_numerical}
\end{equation}
\end{enumerate}
The numerical implementation of the rim-lamella model involves a derivative $\mathd U/\mathd t=(\cdots)/V$, where $V$ is small at early times.  This makes the equations stiff.  For this reason, we have used a high-order accurate numerical solver (ODE89 in Matlab).  Results are shown in Figure~\ref{fig:scaling1}.
Beyond $\myWe=10^2$, the contribution $2a$ in the formula $\mathcal{R}\approx R+2a$ makes little or no difference to the final result.  All numerical results show the clear $\myWe^{1/2}$ scaling from $\myWe=10^2$ to $\myWe=10^6$.  Naturally, the theoretical lower bound~\eqref{eq:R_max_bound_XXX} under-estimates the spreading radius.  At high $\myWe$, the under-estimation is seen from Figure~\ref{fig:scaling1} to be a simple multiplicative factor, whose value (correct to four significant figures) can be inferred from Figure~\ref{fig:scaling1} to be $0.5458$.  

In Figure~\ref{fig:scaling1} we also show the estimate for $R_{max}$ based on perturbation theory (Equation~\eqref{eq:Rmax_upper0}).   Both the rigorous lower bound for $R_{max}$ and the perturbation-theory inspired upper bound for $R_{max}$ possess $\mathrm{We}^{1/2}$ scaling at high Weber number.  Hence, by a `sandwich result', the value of $R_{max}$ coming from the full rim-lamella model must also possess the $\mathrm{We}^{1/2}$ scaling, at sufficiently high Weber number. 
%

\begin{figure}[htb]
	\centering
		\includegraphics[width=0.5\textwidth]{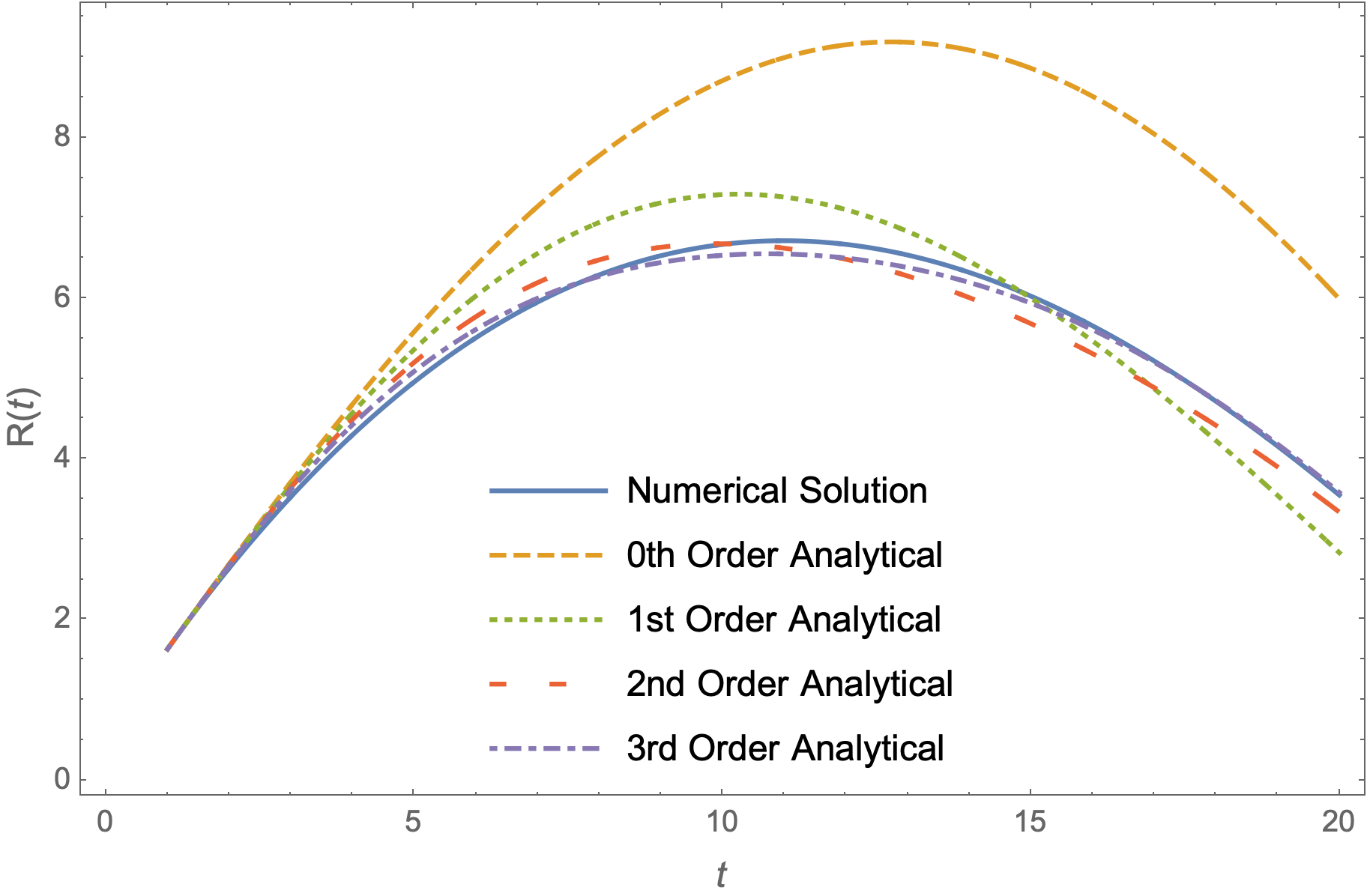}
		\caption{Comparison between the perturbation theory and the numerical solutions of full rim-lamella model.  Parameter values $\myWe=160$ and $\thetadyn=2\pi/3$.}
	\label{fig:fig1_mdb}
\end{figure}

Finally, to further validate  the perturbation theory in Section~\ref{sec:perturbation}, we compare the numerical solutions of full rim-lamella model  (again using the initial conditions~\eqref{eq:ICs_numerical}) with results from the perturbation theory.  Sample results are shown in Figure~\ref{fig:fig1_mdb}.  Using four terms in the expansion, the perturbation theory gives an extremely accurate approximation of $R(t)$.  Using only the zeroth-order term gives an over-estimate of $R(t)$ and hence, an over-estimation of $\max[R(t)]$, which justifies the use the lowest-order theory as an upper bound in Figure~\ref{fig:scaling1}.

\subsection{Comparison between Numerical Solutions of the Rim-Lamella Model and Experiments / DNS}

For the inviscid limit, care is needed in identifying appropriate studies (whether experiments or DNS)  to test the accuracy of the predicted results of the rim-lamella model.  The reason is that in experiments, the no-slip condition cannot be avoided, meaning that a viscous boundary layer forms between the surface and the impacting droplet.  The  boundary layer provides an effective mechanism to arrest the spreading, and induces a Reynolds-number dependent scaling behaviour for $\maxR/R_0$.  The correlation by Roisman~\cite{roisman2009inertia}, $\maxR/R_0\sim a \mathrm{Re}^{1/5}-b \mathrm{We}^{-1/2}\mathrm{Re}^{2/5}$, fits a wide range of experimental conditions.  Here, $a$ and $b$ are constants (for $\mathrm{Re}=U_0 (2R_0)\rho/\mu$ and $\mathrm{We}=\rho U_0^2 (2R_0)/\sigma$ the values are $a = 0.87$ and $b = 0.40$).  However, there are some particular special cases in which the purely inviscid scaling, $\maxR/R_0\sim \mathrm{We}^{1/2}$ manifests itself. For  DNS, this involves simulations where a free-slip boundary condition is applied at the wall~\cite{wildeman2016spreading,wang2022experimental}.  For experiments, one way in which the inviscid scaling can be made manifest is to create a head-on collision of two droplets~\cite{willis2003binary}.  This system of two colliding droplets has a mirror symmetry which is equivalent to a free-slip condition and a contact angle of $\pi/2$.   Another set of experiments attempts to realise spreading under a free-slip boundary condition by treating the impact surface~\cite{wang2022experimental}.   We compare numerical solutions of the rim-lamella model with both experiments and numerical simulations in what follows.

\begin{figure}[htb]
	\centering
		\includegraphics[width=0.5\textwidth]{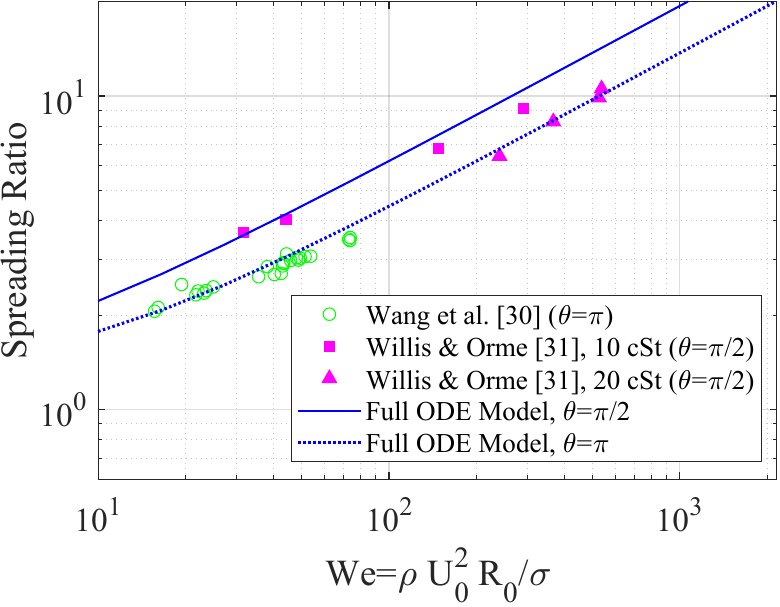}
		\caption{Spreading ratio: Comparison between model and experiments.  Each dataset has a constant contact angle $\theta$.  The (effective) constant contact angle recommended by Wang et al. for their experiments is $\theta=\pi$.~\cite{wang2022experimental}}
	\label{fig:scaling_expts}
\end{figure}
The first set of comparisons concerns the model (full numerical solution) and experiments (Figure~\ref{fig:scaling_expts}).  There is excellent agreement between the model and the experiments of Wang et al.~\cite{wang2022experimental}.  There is also excellent agreement between the model and the experiments of Willis and Orme for 
the $10\,\mathrm{cSt}$ fluid~\cite{willis2003binary}.  Some viscosity-dependence is in evidence in the case of the $20\,\mathrm{cSt}$ fluid as the experimentally-measured maximum spreading ratio is lower than the (inviscid) theoretical prediction in that case.  
\begin{figure}[htb]
	\centering
		\includegraphics[width=0.5\textwidth]{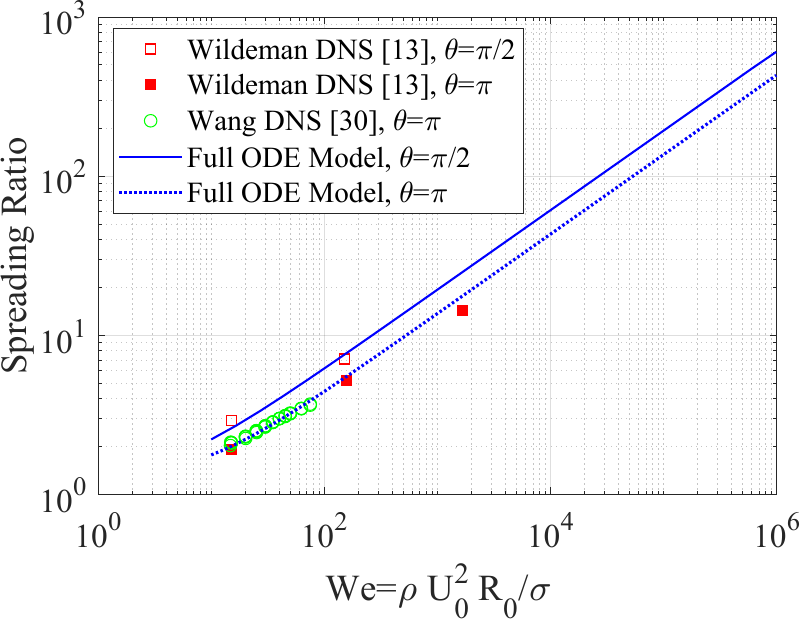}
		\caption{Spreading ratio: Comparison between  model and DNS.  Each dataset has a constant contact angle $\theta$.  }
	\label{fig:scaling_dns}
\end{figure}
A second set of comparisons concerns the model (full numerical solution) and DNS (Figure~\ref{fig:scaling_dns}).  The DNS results are carried out using the free-slip boundary condition, meaning that no boundary layer is formed and hence, spreading arrest due to viscous dissipation does not play a role.  There is excellent agreement between the model and the DNS.
Overall, all results show the clear $\mathrm{We}^{1/2}$ scaling, and close agreement between the experimental / DNS results and the model predictions.   

\begin{figure}[htb]
	\centering
		\includegraphics[width=0.5\textwidth]{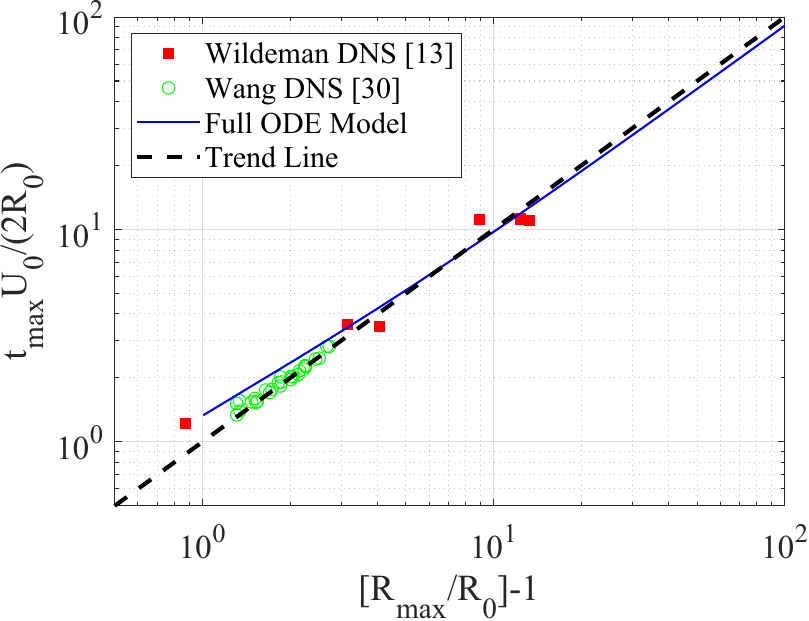}
		\caption{Plot showing the relationship between $t_{max}$ and $(\maxR/R_0)-1$.  The data from Wildeman et al. correspond to all Reynolds numbers considered therein ($\mathrm{Re}=100,500,1000$).  The data from  Wang et al. correspond to $\mathrm{Re}\geq 500$. The trend line is a straight line passing through the origin with slope $45^\circ$.  All datasets use the same constant contact angle $\theta=\pi$.}
	\label{fig:tmax_compare}
\end{figure}

A final comparison concerns the dependence of $\maxR/R_0$ on $t_{max}$, the time at which maximum spreading occurs.  A comparison between the model predictions and the available DNS data is shown in Figure~\ref{fig:tmax_compare}.  As $\tau$ is not known from the model (only $\tau+t_0$ is known), this is estimated as $\tau=R_0/U_0$.  Because of the way in which the time is measured in the model, the value of $\tau$ needs to be added to the time of maximum spreading in the model, to produce the overall value of $t_{max}$.  Again, there is good agreement between the model and the DNS, as evidenced in Figure~\ref{fig:tmax_compare}.
 At large values of $\myWe$, these results all imply $R_{max}/(U_0 t_{max})=0.5$, independent of Weber number.  In this context, the estimate $R_{max}/(U_0 t_{max})\approx 4/(\pi \sqrt{3})\approx 0.7351$ derived from perturbation theory (Equation~\eqref{eq:Rmax_upper1}) is a good approximation.

We emphasize that a similar exercise has already been carried out by Wildeman~\cite{wildeman2016spreading}, with the purpose of validating a correlation for $\maxR/R_0$ based on an energy balance; the point of departure of the present work is to use a momentum balance (the rim-lamella model).  Mathematically, both approaches are equivalent: the present agreement between the experiments and the DNS is an (indirect) validation of this equivalence.  Finally, we emphasize that the experiments by Wang et al.~\cite{wang2022experimental} allow for much smaller values of Weber number than those investigated herein.  The value of the maximum spreading radius predicted by the model and measured in the experiments is again close in these cases (not shown).  However, this should be regarded as fortuitous, as a rim-lamella structure is not in evidence at low Weber numbers ($\rho U_0^2 (2R_0)/\sigma\sim 30$, \cite{wildeman2016spreading}).

\section{Discussion and Conclusions}
\label{sec:conclusions}

Summarizing, we have revisited the problem of droplet impact and droplet spread on a smooth surface in the case of an ideal inviscid fluid.  This problem is well studied in the literature, and there is a canonical model which describes the maximum spreading radius, due to Roisman et al.~\cite{roisman2002normal}.  We have revisited this model with two aims in mind:
\begin{enumerate}[noitemsep]
\item To perform a theoretical analysis of the ODE system arising from the rim-lamella model;
\item To demonstrate self-consistency of the rim-lamella model.
\end{enumerate}
This article fulfils these aims.  Using Gronwall's Inequality, our theoretical analysis of the ODE system has yielded a lower bound for the maximum spreading radius such that $\max[R(t)] \geq (\mathrm{Const.})\times \myWe^{1/2}$ at sufficiently large Weber number.  Using perturbation theory, we have further shown that $\max[R(t)] \leq (\mathrm{Const.})\times \myWe^{1/2}$.  Hence, by a sandwich result, we have demonstrated -- from first principles -- the classical scaling behaviour $\max[R(t)]\sim \mathrm{We}^{1/2}$, valid at sufficiently high Weber number.   
Using the same theoretical analysis, we have demonstrated  rigorously that the  rim-lamella model is self-consistent: once a rim forms, its height will invariably exceed that of the lamella.  

To determine the sharpness of our estimates, we have introduced a physics-inspired methodology to approximate the initial conditions for the rim-lamella model.  Our approach is based on other studies in the literature, notably~\cite{roisman2002normal,fedorchenko2004some,eggers2010drop}.
%
%
We have used our approximations for the initial conditions for the rim-lamella model -- together with high-accuracy numerical solutions of same  -- to compare the model results for
 the maximum spreading radius with the available DNS and experimental evidence, and excellent agreement between the various approaches is obtained.

The present work has focused on the ideal inviscid case, for which there is no viscous boundary layer and hence, no mechanism for viscosity to arrest the spreading.  The reason for considering the ideal inviscid case is to introduce a suite of theoretical methods for characterizing the rim-lamella model.  We have sought to showcase in particular the application of the theory of estimates to the problem of droplet spreading, as this theory has been used  successfully to generate key results in other problems in Fluid Dynamics -- notably, hydrodynamic stability theory, regularity theory for the Navier--Stokes Equations, mixing, and in the case of the mathematical analysis of the Thin-Film Equation.  
 These techniques may be employed in future to the droplet-spreading for fluids with non-zero viscosity, in which case the present theoretical methods might yield rigorous estimates for the scaling behaviour for the maximum spreading radius with both Weber number and Reynolds number.

\subsection*{Acknowlegements}

This work has been produced as part of ongoing work within the ThermaSMART network.
The ThermaSMART network has received funding from the European Union's Horizon 2020
research and innovation programme under the Marie Sklodowska--Curie grant agreement No.
778104.

\appendix

\section{Further Analysis of the Rim-Lamella Model}
\label{sec:AppA}

In this Appendix, we present a theoretical analysis of the ODE model~\eqref{eq:roisman}.  Specifically, we prove a sandwich result for $\Delta=u_0-U$:
\begin{equation}
0< \Delta(\tau)\left(\frac{\tau+t_0}{t+t_0}\right)           \leq       \Delta \leq c(t),
\label{eq:sandwich1}
\end{equation}
provided the initial conditions satisfy $0\leq \Delta(\tau)\leq c(\tau)$.  
The result~\eqref{eq:sandwich1} is valid for all $t\in [0,t_{max}]$, where $t_{max}$ is the turnover time such that $U(t_{max})=0$.  The result further relies on the initial conditions
\begin{equation}
U_{init}>0,\qquad \Delta(\tau)>0,\qquad \frac{3\eta}{2 U_0^2} \left[\frac{R_{init}}{\tau+t_0}+\Delta(\tau)\right]^2<1.
\label{eq:app:bds}
\end{equation}
The appendix should be be read in conjunction with Section~\ref{sec:consistent}.

We recall first the equation for $\mathd U/\mathd t$:
\begin{equation}
\frac{\mathd U}{\mathd t}=\frac{2\pi R}{ V}\left[h\Delta^2-{\frac{\sigma}{\rho}(1-\cos\thetadyn)}\right].
\end{equation}
We identify:
\begin{equation}
c(t)=\sqrt{\frac{\sigma(1-\cos\thetadyn)}{\rho h}}=\sqrt{\frac{\sigma(1-\cos\thetadyn)}{\rho R_0 \eta}} (U_0/R_0)(t+t_0)\mathe^{(3\eta/8U_0^2)[R/(t+t_0)]^2}.
\end{equation}
Hence,
\begin{equation}
V\frac{\mathd U}{\mathd t}=\frac{2\pi R h }{V}\left(\Delta^2-c^2\right).
\end{equation}
We formulate an equation for $\Delta$ by direct differentiation:
\begin{equation}
\frac{\mathd \Delta}{\mathd t}+\frac{\Delta}{t+t_0}=-\frac{\mathd U}{\mathd t}.
\end{equation}
The right-hand side can be identified as $-(2\pi Rh/V)\left(\Delta-c\right)\left(\Delta+c\right)$, hence:
%
%
\begin{equation}
\frac{\mathd \Delta}{\mathd t}+\frac{\Delta}{t+t_0}=-\frac{2\pi R h}{V}\left(\Delta-c\right)\left(\Delta+c\right).
\label{eq:factor}
\end{equation}
Thus,
\begin{equation}
\frac{\mathd}{\mathd t}(\Delta-c)=-\frac{\Delta}{t+t_0}-\frac{2\pi R h}{V}\left(\Delta-c\right)\left(\Delta+c\right)-\frac{\mathd c}{\mathd t}.
\end{equation}
By direct differentiation of $c(t)$:
\begin{equation}
\frac{\mathd c}{\mathd t}=\frac{c}{t+t_0}\left( 1-\frac{3\eta}{4U_0^2} \frac{R}{t+t_0}\Delta\right).
\end{equation}
Introduce $Y=\Delta-c$.  We have:
\begin{equation}
\frac{\mathd Y}{\mathd t}=-\frac{Y}{t+t_0}-\frac{2\pi R h}{V}Y\left(Y+2c\right)-\frac{2c}{t+t_0}\left( 1-\frac{3\eta}{8U_0^2} \frac{R}{t+t_0}\Delta\right).
\label{eq:ineq0}
\end{equation}
We have:
\begin{eqnarray*}
\frac{3\eta}{8U_0^2}\frac{R_{init}}{\tau+t_0}\Delta(\tau)&=& \frac{3\eta}{8U_0^2}\frac{R_{init}}{\tau+t_0} \left( \frac{R_{init}}{\tau+t_0}-U_{init}\right),\\
&\leq & \frac{3\eta}{8U_0^2}\left(\frac{R_{init}}{\tau+t_0}\right)^2,\\
&\leq & \frac{3\eta}{8U_0^2}\left[\frac{R_{init}}{\tau+t_0}+\Delta(\tau)\right]^2,\\
&\stackrel{\text{I.C.s}}{<}& 1.
\end{eqnarray*}
Thus,
\[
\frac{3\eta}{8U_0^2}\frac{R_{init}}{\tau+t_0}\Delta(\tau)<1,
\]
and by continuity, there is an interval $I=[\tau,t_*)$ such that
\[
\frac{3\eta}{8U_0^2} \frac{R(t)}{t+t_0}\Delta<1,\qquad t\in I.
\]
On this interval, $1-(3\eta/8U_0^2)[R/(t+t_0)]\Delta >0$.  Equation~\eqref{eq:ineq0} can then be turned into a differential inequality:
\begin{equation}
\frac{\mathd Y}{\mathd t}+\left(\frac{1}{t+t_0}+\frac{4\pi R c h}{V}\right)Y< 0,\qquad t\in I.
\label{eq:ineq1}
\end{equation}
This is a linear first-order differential inequality; the integrating factor is $\mu=(t+t_0)\mathrm{e}^{\int_\tau^{t}(4\pi R ch/V)\mathd t}$.
%
We apply Gronwall's Lemma to obtain:
\[
Y(t)(t+\tau) \leq Y(\tau)(\tau+t_0)\mathrm{e}^{-\int_\tau^{t}(4\pi R ch/V)\mathd t}\leq Y(\tau)(\tau+t_0),\qquad t\in I.
\]
Hence,
\[
\Delta- c\leq  [\Delta(\tau)-c(\tau)]\left(\frac{\tau+t_0}{t+t_0}\right),\qquad t\in I.
\]
We use the given initial condition, $0\leq \Delta(\tau)\leq c(\tau)$, such that 
\[
\Delta - c \leq 0,\qquad t\in I.
\]
at later times.  Hence, $\Delta \leq c$ for $t\in I$.  Going back to Equation~\eqref{eq:factor}, we see that
\[
\frac{\mathd \Delta}{\mathd t}+\frac{\Delta}{t+t_0}\geq 0,\qquad t\in I.
\]
A second application of Gronwall's inequality yields:
\[
\Delta(t)(t+t_0)\geq \Delta(\tau)(\tau+t_0),\qquad t\in I.
\]
Thus,
\begin{equation}
0\leq \Delta(\tau)\left(\frac{\tau+t_0}{t+t_0}\right)           \leq       \Delta \leq c(t).
\end{equation}
Hence also, $\Delta$ is bounded away from zero and is therefore positive for all $t\in I$.

Next, we look at $\Delta=R/(t+t_0)-\mathd R/\mathd t$.  Hence,
\[
\frac{\mathd R}{\mathd t}-\frac{R}{t+t_0}=-\Delta \leq -\Delta(\tau)\frac{\tau+t_0}{t+t_0}.
\]
A third and final application of Gronwall's inequality gives:
\begin{equation}
R(t)\leq \left[\frac{R_{init}}{\tau+t_0}-\Delta(\tau)\right](t+t_0)+\Delta(\tau)(\tau+t_0).
\end{equation}
Hence,
\begin{equation}
\frac{R(t)}{t+t_0}\leq \frac{R_{init}}{\tau+t_0}+\Delta(\tau).
\label{eq:app:Rtfin}
\end{equation}
Finally, we look at 
\begin{equation}
\chi=\frac{3\eta}{8U_0^2}\frac{R}{t+t_0}\Delta=\frac{3\eta}{8U_0^2}\frac{R}{t+t_0}\left(\frac{R}{t+t_0} - U\right).
\end{equation}
Let $t_{max}$ denote the point of maximum spreading, such that $U(t_{max})=0$.    We look at the interval $J=[\tau,\min(t_*,t_{max}))$.   In view of the initial conditions~\eqref{eq:app:bds}, $U(t)>0$ for $t\in J$.  Hence,
\begin{eqnarray*}
\chi &=& \frac{3\eta}{8U_0^2}\frac{R}{t+t_0}\left(\frac{R}{t+t_0} - U\right),\\
     &\leq &\frac{3\eta}{8U_0^2 }\left(\frac{R}{t+t_0}\right)^2,\qquad t\in J,\\
		 &\stackrel{\text{Eq.~\eqref{eq:app:Rtfin}}}{\leq} &\frac{3\eta}{8U_0^2} \left[\frac{R_{init}}{\tau+t_0}+\Delta(\tau)\right]^2,\\
		 &\stackrel{\text{I.C.s}}{<}& 1.
\end{eqnarray*}
It therefore follows that the right-hand side of Equation~\eqref{eq:ineq0} is negative or zero for all $t\in [\tau,t_{max})$ and hence, all of the foregoing results can be extended from the interval $I$ to the interval $[\tau,t_{max}]$.  In particular,
\begin{subequations}
\begin{eqnarray}
0&\leq & \Delta(\tau)\left(\frac{\tau+t_0}{t+t_0}\right)           \leq       \Delta \leq c(t),\\
R(t)&\leq &\left[\frac{R_{init}}{\tau+t_0}-\Delta(\tau)\right](t+t_0)+\Delta(\tau)(\tau+t_0),
\end{eqnarray}
\end{subequations}
for all $t\in [0,t_{max}]$.

\section{Collapse of the Rim-Lamella Model into a single equation}
\label{app:collapse}

\renewcommand{\theequation}{\thesection.\arabic{equation}}

In this Appendix, we show how the rim-lamella model~\eqref{eq:roisman} can be collapsed into a single equation.  In particular, we consider Equation~\eqref{eq:mass} and~\eqref{eq:Deltadef}, 
which we rewrite again here in terms of the variables $V$, $\Delta$ and $R$:
\begin{eqnarray}
\label{eq:V_mdb}
\frac{\mathd V}{\mathd t}&=& 2\pi R h(R,t) \Delta \,
,\\
\label{eq:Delta_mdb}
\frac{\mathd \Delta}{\mathd t}+\frac{\Delta}{t+t_0}&=&-\frac{2\pi  R h(R,t)}{V} \big\{\Delta^2-[c(t)]^2\big\}\,,
\end{eqnarray}
Here, as before, $c(t)$ denotes the characteristic speed.  

The starting-point is to multiply Equation~\eqref{eq:V_mdb} by $\Delta$ and equation (\ref{eq:Delta_mdb}) by $V$.  We add the results and get:
$$\frac{\mathd (V \Delta)}{\mathd t}+\frac{V \Delta}{t+t_0} = 2 \pi \frac{\sigma}{\rho} (1-\cos\thetadyn) R\,,$$
We multiply this equation by $(t+t_0)$ in order to write the LHS as a derivative of a product:
$$\frac{\mathd ((t+t_0) V \Delta)}{\mathd t} = 2 \pi \frac{\sigma}{\rho} (1-\cos\thetadyn) (t+t_0) R\,.$$
Finally we use equation (\ref{eq:V_mdb}) to eliminate $\Delta$ in the above equation, giving
\begin{equation}
\label{eq:ODE1_mdb}
\frac{\mathd}{\mathd t}\left[\frac{(t+t_0)}{2\pi R h(R,t)} V \frac{\mathd V}{\mathd t}\right] = 2 \pi \frac{\sigma}{\rho} (1-\cos\thetadyn) (t+t_0) R\,.
\end{equation}
From here on, to simplify the analysis we use the `remote asymptotic solution', Equation~\eqref{eq:RAS}, for the drop height:
$$h(R,t) = h_{init}\left(\frac{\tau+t_0}{t+t_0}\right)^2\,,$$
so equation (\ref{eq:ODE1_mdb}) reads now
$$\frac{\mathd}{\mathd t}\left[\frac{(t+t_0)^3}{2 \pi (\tau+t_0)^2 h_{init}  R} V \frac{\mathd V}{\mathd t}\right] = 2 \pi \frac{\sigma}{\rho} (1-\cos\thetadyn) (t+t_0) R\,.
$$
In this approximation the total mass conservation reduces to
$\pi h(R,t) R^2 + V =  V_{tot}$ (constant),
which allows us to eliminate $V = V_{tot}-\pi h_{init}(\tau+t_0)^2\left(\frac{R}{t+t_0}\right)^2$ in the above ODE, giving
$$\frac{\mathd}{\mathd t}\bigg\{\frac{(t+t_0)^3}{2 R} \left[V_{tot}-\pi h_{init}(\tau+t_0)^2\left(\frac{R}{t+t_0}\right)^2\right] \frac{\mathd \left(\frac{R}{t+t_0}\right)^2 }{\mathd t}\bigg\} = - 2 \pi   \frac{\sigma}{\rho} (1-\cos\thetadyn) (t+t_0) R\,,
$$
which can be further rearranged to 
$$\frac{\mathd}{\mathd t}\bigg\{(t+t_0)^2 \left[V_{tot}-\pi h_{init}(\tau+t_0)^2\left(\frac{R}{t+t_0}\right)^2\right] \frac{\mathd \left(\frac{R}{t+t_0}\right) }{\mathd t}\bigg\} = - 2 \pi   \frac{\sigma}{\rho} (1-\cos\thetadyn) (t+t_0) R\,,
$$
and finally to
\begin{equation*}
\frac{\mathd}{\mathd t}\bigg\{(t+t_0)^2 \frac{\mathd}{\mathd t} \left[\frac{R}{t+t_0}-\frac{(\tau+t_0)^2}{3 R_{init}^2}\left(\frac{R}{t+t_0}\right)^3\right]\bigg\} = - \frac{2 c(\tau)^2}{R_{init}^2} (t+t_0)^2 \frac{R}{t+t_0},
\end{equation*}
which is Equation~\eqref{eq:ODE_R_mdb} in the main paper.

\end{document}